\documentclass[aps,prd,twocolumn,showpacs,superscriptaddress,letterpaper,amsmath,amssymb]{revtex4}

\usepackage[T1]{fontenc} 
\usepackage{graphicx}
\usepackage{graphics}
\usepackage[mathscr]{eucal}
\usepackage{amssymb}
\usepackage{amsmath}
\usepackage{xspace}
\usepackage{listings}
\usepackage{soul}
\usepackage{ulem}
\usepackage{color}
\usepackage[dvipsnames]{xcolor}
\usepackage{ulem}
\usepackage{bm}         
\usepackage{array}
\usepackage{multirow}   
\usepackage{booktabs}   
\usepackage{mleftright}
\usepackage[utf8]{inputenc}
\newcommand{\vev}[0]{VEV\xspace}
\newcommand{\vevs}[0]{VEVs\xspace}

\newcommand{\tr}{{\rm Tr}}

\renewcommand{\i}{\mathrm i}
\newcommand*{\Scale}[2][4]{\scalebox{#1}{$#2$}}%

\newcommand{\U}[1]{\mathrm{U}(1)_{\mathrm{#1}}}			
\newcommand{\SU}[2]{\mathrm{SU}(#1)_{\mathrm{#2}}}		

\newcommand{\TuL}[3]{\big(T_{\mathrm{L}}^{#1} \big)_{#2}^{#3}}

\newcommand{\LLR}[3]{\big(\bm{L}^{ #1} \big)^{ #2 }_{ #3 }}
\newcommand{\QL}[3]{\big(\bm{Q}_{\mathrm{L}}^{ #1} \big)^{ #2 }_{ #3 }}
\newcommand{\QR}[3]{\big(\bm{Q}_{\mathrm{R}}^{ #1} \big)^{ #2 }_{ #3 }}

\newcommand{\LLRs}[3]{\big(L^{ #1} \big)^{ #2 }_{ #3 }}

\newcommand{\SLLR}[3]{\big(\tilde{L}^{ #1} \big)^{ #2 }_{ #3 }}

\newcommand{\SLLRs}[3]{\big(\tilde{L}^*_{ #1} \big)^{ #3 }_{ #2 }}

\begin{document}

\title{Reviving trinification models through an $\mathrm{E}_6$-extended supersymmetric GUT}

\author{Jos{\'e}~E.~Camargo-Molina}
\affiliation{Department of Astronomy and Theoretical Physics, Lund University, 221 00 Lund, Sweden}

\author{Ant{\'o}nio~P.~Morais}
\affiliation{Department of Astronomy and Theoretical Physics, Lund University, 221 00 Lund, Sweden}
\affiliation{Departamento de F\'isica, Universidade de Aveiro and CIDMA, Campus de Santiago, 3810-183 Aveiro, Portugal}

\author{Astrid~Ordell}
\affiliation{Department of Astronomy and Theoretical Physics, Lund University, 221 00 Lund, Sweden}

\author{Roman~Pasechnik}
\affiliation{Department of Astronomy and Theoretical Physics, Lund University, 221 00 Lund, Sweden}

\author{Marco~O.~P.~Sampaio}
\affiliation{Departamento de F\'isica, Universidade de Aveiro and CIDMA, Campus de Santiago, 3810-183 Aveiro, Portugal}

\author{Jonas~Wess\'en}
\affiliation{Department of Astronomy and Theoretical Physics, Lund University, 221 00 Lund, Sweden}

\begin{abstract}
We present a supersymmetric (SUSY) model based on trinification $[\SU{3}{}]^3$ and family $\SU{3}{F}$ symmetries embedded 
into a maximal subgroup of $\mathrm{E}_8$, where the sectors of light Higgs bosons and leptons are unified into a single chiral supermultiplet. 
The common origin of gauge trinification and of the family symmetry from $\mathrm{E}_8$ separates the model from other trinification-based GUTs, 
as it protects, in particular, the Standard Model fermions from gaining mass until the electroweak symmetry is broken. Furthermore, it allows us to break 
the trinification symmetry via vacuum expectation values in $\SU{3}{}$-adjoint scalars down to a left-right symmetric theory. Simultaneously, it ensures the unification of 
the gauge and Yukawa couplings as well as proton stability. Although the low-energy regime (e.g.~mass hierarchies in the scalar sector determined 
by a soft SUSY-breaking mechanism) is yet to be established, these features are one key to revive the once very popular trinification-based GUTs.
\end{abstract}


\maketitle

\section{Introduction}

Finding a compelling theory for the unification of the fundamental interactions that is capable of reproducing known 
features of the Standard Model (SM) has been a major goal of the theoretical physics community. Popular SM extensions are supersymmetric (SUSY) 
grand unified theories (GUTs) based on simple Lie groups such as e.g.~$\SU{5}{}$ \cite{Georgi:1974sy}, $\mbox{SO}(10)$ \cite{Fritzsch:1974nn}, 
$\mathrm{E}_6$ \cite{Gursey:1975ki}, and $\mathrm{E}_7$ \cite{Gursey:1976dn}. However, many of the existing GUTs typically suffer from various 
issues with, e.g.,~proton stability, fine-tuning, and hierarchies in parameters such as fermion masses and mixings lacking a fundamental explanation, 
as well as with inconceivably complicated parameter spaces severely reducing their predictive power.

GUTs inspired by $\mathrm{E_6}$ are becoming increasingly popular due to their rich phenomenology and their many attractive properties 
(see, e.g.,~Refs.~\cite{King:2005my,King:2007uj,Braam:2010sy,Nevzorov:2012hs}). One such GUT scenario based upon a maximal 
rank-6 subgroup $[\SU{3}{}]^3 \subset \mathrm{E}_6$ and known as gauge trinification (T-GUT) was initially proposed by Glashow et al in 1984 \cite{original}. 
The trinification symmetry is typically identified as a left-right-color product group, {\it i.e.},~$[\SU{3}{}]^3\equiv \SU{3}{L} \times  \SU{3}{R} \times  
\SU{3}{C}$, and is supplemented by a cyclic permutation symmetry $\mathbb{Z}_3$ forcing the gauge couplings to unify, 
{\it i.e.}~$g_{\mathrm{U}}\equiv g_{\mathrm{L}}=g_{\mathrm{R}}=g_{\mathrm{C}}$. One of the appealing features of T-GUT models is that 
all the matter fields, which belong to bitriplet representations (reps) of the trinification symmetry,
\begin{eqnarray}
\Scale[0.95]{\LLR{i}{l}{r} =\begin{pmatrix}
\bm{H}_{11} & \bm{H}_{12} & \bm{e}_{\mathrm{L}}\\
\bm{H}_{21} & \bm{H}_{22} & \bm{\nu}_{\mathrm{L}}\\
\bm{e}_{\mathrm{R}}^{c} & \bm{\nu}_{\mathrm{R}}^{c} & \bm{\phi}
\end{pmatrix}^{i},} \hspace{2mm}
\begin{aligned}
&\Scale[0.95]{\QL{i}{x}{l}=\begin{pmatrix}\bm{u}_{\mathrm{L}}^x & \bm{d}_{\mathrm{L}}^x & \bm{D}_{\mathrm{L}}^x
\end{pmatrix}^{i}}, \label{eq:tri-triplets}
\\
&\Scale[0.95]{\QR{i}{r}{x}=\begin{pmatrix}\bm{u}_{\mathrm{R}x}^c & \bm{d}_{\mathrm{R}x}^c & \bm{D}_{\mathrm{R}x}^c
\end{pmatrix}^{\top\;\;i}},
\end{aligned}
\end{eqnarray}
can be embedded into three $\bm{27}$-plets of $\mathrm{E}_6$ as $\bm{27}^i \to (\bm{3},\bm{\bar{3}},\bm{1})^i\oplus 
(\bm{\bar{3}},\bm{1},\bm{3})^i\oplus (\bm{1},\bm{3},\bm{\bar{3}})^i$. Here, the left, right, and color $\SU{3}{}$ 
indices are $l,\,r,$ and $x$, respectively, while the generations are labeled by an index $i$ (for an alternative realization containing the trinified gauge symmetry $[\SU{3}{}]^3$, 
see Refs.~\cite{Dias:2010vt,Reig:2016tuk}). Some T-GUT versions claim to preserve baryon number naturally \cite{Babu:1985gi, Sayre:2006ma} but can also 
be engineered to account for the baryon-antibaryon asymmetry in the Universe through heavy Higgs decays at one loop \cite{He:1986cs}. They can, in principle, accommodate 
any quark and lepton masses and mixing angles \cite{Babu:1985gi} while neutrino masses can be generated by, e.g.,~a radiative \cite{Sayre:2006ma} or an inverse 
\cite{Cauet:2010ng} see-saw mechanism. However, despite some progress in recent years, the T-GUT scenarios remain among the least explored extensions of the SM. 
One of the major theoretical challenges in building the SUSY-based T-GUTs is finding a stable vacuum with spontaneously broken gauge trinification while keeping 
a low number of free parameters at the GUT scale.

In order to avoid GUT-scale lepton masses, previous realizations of T-GUTs introduced either additional unmotivated Higgs multiplets
\cite{Babu:1985gi,Wang:1992hu,Dvali:1994wj,Maekawa:2002qv,Willenbrock:2003ca,Carone:2005ha,Sayre:2006ma,Cauet:2010ng,
Stech:2012zr,Stech:2014tla,Hetzel:2015bla,Pelaggi:2015kna}, whose vacuum expectation values (\vevs) provide a consistent spontaneous symmetry breaking (SSB) 
of trinification down to the SM gauge symmetry, or higher-dimensional operators \cite{Nath:1988xn,Dvali:1994wj,Maekawa:2002qv,Carone:2005ha,Cauet:2010ng}. 
Such constructions may, however, result in severe phenomenological contradictions with proton stability \cite{Babu:1985gi,Maekawa:2002qv,Sayre:2006ma} and 
too many unobserved low-scale signatures \cite{original,Nath:1988xn,Dvali:1994wj,Stech:2014tla,Hetzel:2015bla,Pelaggi:2015knk}. As a consequence, a large number of free 
Yukawa parameters in the superpotential has to be highly fine-tuned in order to reproduce the SM mass hierarchies \cite{Sayre:2006ma}. A proper renormalization group (RG) analysis 
of a high-scale SUSY model containing a few hundreds of particles and couplings and accounting for several SSB scales down to the effective low-energy SM-like theory 
remains barely feasible in practice. Thus, deriving even basic features of the SM (such as fermion mass/mixing hierarchies and Higgs sector properties) as a low-energy 
effective field theory (EFT) limit of a T-GUT remains a big unsolved problem (for more details, see, e.g.,~Ref.~\cite{Hetzel:2015cca} and references therein). 

In this paper, we propose a new way to resolve the problem of GUT-scale masses of the SM leptons inspired by an embedding of the trinification 
$[\SU{3}{}]^3\subset \mathrm{E}_6$ and family $\SU{3}{F}$ symmetries into the maximal exceptional symmetry group $\mathrm{E}_8$. A common origin 
of family symmetry and SM gauge symmetries from $[\SU{3}{}]^3\times \SU{3}{F}\subset\mathrm{E}_8$ implies that, in particular, the light Higgs 
and lepton sectors originate from the same (tritriplet) rep of $[\SU{3}{}]^3\times \SU{3}{F}$. Having such light Higgs-lepton unification in 
the $\mathrm{E}_6$-extended theory (inspired by $\mathrm{E}_8$) leads to a complete unification of quark and lepton Yukawa couplings for all three generations (as well as 
the quartic interactions of the scalar potential) at the trinification-breaking scale. This is at variance with popular $\mathrm{SO}(10)$ and 
Pati-Salam models where the unification of Yukawa couplings is restricted to the third family \cite{Blazek:2001sb, Baer:2001yy, Anandakrishnan:2013cwa, 
Blazek:2002ta, Tobe:2003bc, Baer:2009ie, Badziak:2011wm, Anandakrishnan:2012tj, Joshipura:2012sr, Anandakrishnan:2013nca, Anandakrishnan:2014nea, 
Badziak:2013eda, Ajaib:2013zha}. Such a distinct feature of the high-scale model dramatically reduces its parameter space making its complete analysis 
computationally simple, at least, at tree level. We have found that the proposed $\mathrm{E}_6$-extended T-GUT model gives rise to an effective 
left-right (LR) symmetric theory with specific properties. The remnant $\SU{3}{F}$ family symmetry reduces the number of allowed 
terms in the LR-symmetric EFT, simplifying its matching procedure with the high-scale theory and making its RG flow analysis technically feasible.
A consistent match of the LR-symmetric EFT with the SM at low scale would then strongly constrain the hierarchies in the soft SUSY-breaking 
sector offering new possibilities for studies of the SUSY breaking in $\mathrm{E}_6$-based theories.

\section{$\mathrm{E}_8$-inspired family symmetry}

In earlier work by some of the authors~\cite{Camargo-Molina:2016bwm}, 
it was understood that the SM gauge group can arise dynamically from a non-SUSY T-GUT in a scenario where fermions and scalars belong to the same 
$\mathrm{E}_6$ reps [augmented by a global $\SU{3}{F}$], thus hinting at a possible presence of SUSY at (or beyond) the GUT scale. In particular, 
the color-neutral scalars $\tilde{L}$ (containing the Higgs scalars) and fermions $L$ (containing the SM leptons and right-handed neutrinos) could then be naturally 
considered as components of $\bm{L}$. Here and below, the notations $\tilde{f}$ and $f$ for scalar and fermion components 
of the superfield $\bm{f}$ are used.

Inspired by this observation, the implications of a Higgs-lepton unification in a SUSY T-GUT were explored, with \textit{local} gauge 
trinification $[\SU{3}{}]^3$ and \textit{global} family $\SU{3}{F}$ motivated by a minimal $\mathrm{E}_6$ embedding into $\mathrm{E}_8$ as $\mathrm{E}_6 
\times \SU{3}{F}\subset \mathrm{E}_8$ \cite{Slansky:1981yr,Green:2012oqa}. Indeed, such an $\mathrm{E}_6$-extended trinification model inspired by 
its $\mathrm{E}_8$ embedding can be considered as an approximation to the full gauge $[\SU{3}{}]^3\times \SU{3}{F}\subset \mathrm{E}_8$ theory 
in those regions of parameter space where gauge $\SU{3}{F}$ interactions are suppressed, $g_{\mathrm{F}}\ll g_{\mathrm{U}}$. A special interest 
in $\mathrm{E}_8$-based models originates from string theories where massless sectors are described by the $\mathrm{E}_8\times \mathrm{E}_8^\prime$ 
symmetry~\cite{Candelas:1985en,Green:2012oqa}.

At variance with the non-SUSY model \cite{Camargo-Molina:2016bwm}, incorporating the $\SU{3}{F}$ family symmetry in a SUSY T-GUT model with only 
tritriplets of $[\SU{3}{}]^3\times \SU{3}{F}$ (specified in the first three rows of Table \ref{table:ChiralSuperE8}) leads to a scalar potential containing flat 
directions with color-breaking \vevs. Even with the inclusion of soft breaking terms, such a model at tree level is necessarily inconsistent with the SM at 
low scales. Alternatively, the desired trinification SSB becomes possible in a SUSY T-GUT when relaxing $\SU{3}{F}$. However, this reintroduces GUT-scale 
masses for those SM leptons that are SUSY partners of the Goldstone bosons from $\tilde{L}$, due to terms such as 
$\Scale[0.9]{ -\sqrt{2} g_{\mathrm{U}} \SLLRs{i}{l_1}{r} \TuL{a}{l_2}{l_1} \LLRs{i}{l_2}{r} \lambda^a_{\mathrm L} }$. These terms lead to gaugino-lepton mass 
terms of the order of the T-GUT-breaking \vev $\tilde{L}^i$. Although components in the trinification gaugino fields $\lambda^a_{\mathrm{L,R}}$ could in 
principle build up one generation of the SM leptons, we find such a construction unappealing both due to the reduction of the family symmetry and the 
abandonment of the full Higgs-lepton unification. Besides, the gaugino mass scale in this case would then be unnaturally small for a consistency with 
the SM lepton sector.

This gaugino-lepton mixing indeed posed a big problem for early attempts to consistently unify the Higgs and lepton sectors. However, rather than including 
additional copies of $\bm{L}$, we have found that the leptons are protected from obtaining GUT-scale masses via the inclusion of $\SU{3}{}$ adjoint superfields which, 
together with tritriplets $\bm{L}$, $\bm{Q}_{\mathrm L}$, and $\bm{Q}_{\mathrm R}$, are irreducible representations (irreps) of the $\mathrm{E}_8$ symmetry group.
This novel scenario is in the focus of our further discussion.

\section{Minimal $\mathrm{E}_6$-extended T-GUT model}

The proposed [$\mathbb{Z}_2\times \mathbb{Z}_3$]-symmetric $\mathrm{E}_6$-extended model, where 
the problem of SUSY T-GUT breaking is consistently resolved, preserves all the well-known attractive features of T-GUTs. The chiral superfield content of this model 
transforms as (\textbf{8},\textbf{1}), (\textbf{3},\textbf{27}), and (\textbf{1},\textbf{78}) of $\SU{3}{F}\times \mathrm{E}_6$, where $\SU{3}{F}$ 
is a global family symmetry. This set contains, in addition 
to the lepton and quark superfields $\bm{L}$, $\bm{Q}_{\mathrm L}$ and $\bm{Q}_{\mathrm R}$, chiral supermultiplets in the adjoint rep of $\SU{3}{A}$ (A=L,R,C,F) 
shown in Table~\ref{table:ChiralSuperE8}. The superpotential of this model reads
\begin{eqnarray}
\begin{aligned}
& W = \sum_{\mathrm{A}=\mathrm{L,R,C}}\,\big[\lambda_{\bm{78}} \, d_{abc} \bm{\Delta}_{\mathrm A}^a \bm{\Delta}_{\mathrm A}^b \bm{\Delta}_{\mathrm A}^c + 
\mu_{\bm{78}} \bm{\Delta}_{\mathrm A}^a \bm{\Delta}_{\mathrm A}^a \big]  \label{eq:WE8} \\ 
&+\;\lambda_{\bm{1}} d_{abc} \bm{\Delta}_{\mathrm F}^a \bm{\Delta}_{\mathrm F}^b \bm{\Delta}_{\mathrm F}^c + 
\mu_{\bm{1}} \bm{\Delta}_{\mathrm F}^a \bm{\Delta}_{\mathrm F}^a + 
\lambda_{\bm{27}} \, \varepsilon_{ijk}  \bm{Q}_{\mathrm L}^i \bm{Q}_{\mathrm R}^j \bm{L}^k \,,
\end{aligned}
\end{eqnarray}
where $\lambda_{\bm{27}}$ is the unified quark-lepton Yukawa coupling, the subscript under the couplings denotes the $\mathrm{E}_6$ irreps, 
$d_{a b c} \equiv 2\, \tr \left[ \left\{T_a, T_b \right\} T_c \right]$ are the totally symmetric $\SU{3}{}$ coefficients, 
$\Scale[0.9]{\bm{Q}_{\mathrm L}^i \bm{Q}_{\mathrm R}^j \bm{L}^k \equiv \QL{i}{x}{l} \QR{j}{r}{x} \LLR{k}{l}{r}}$,
and summation over repeated indices is always implied. Furthermore, $\bm{L}$ unifies the light Higgs scalar and lepton sectors while $\bm{Q}_{\mathrm L}$ and 
$\bm{Q}_{\mathrm R}$ contain the SM quarks. In what follows, we refer to this model as the SUSY Higgs-unified trinification (or, shortly, SHUT) model.
\begin{table}[htbp]
\centering
\begin{tabular}{|>{\centering}m{3.1cm}|>{\centering}m{1.15cm}|>{\centering}m{1.15cm}|>{\centering}m{1.15cm}||>{\centering}m{1.15cm}|}
\hline
\small{Superfield} &\small{$\SU{3}{C}$}&\small{$\SU{3}{L}$}&\small{$\SU{3}{R}$}&\small{$\SU{3}{F}$}\tabularnewline
\hline
\small{Lepton} \hfill \small{$\LLR{i}{l}{r}$}&\small{$ \bm{1} $ }&\small{$\bm{3}^l$}&\small{$\bm{\bar{3}}_r$}&\small{$\bm{3}^{i}$}\tabularnewline
\vspace{1mm}
\small{Right-Quark} \hfill \small{$\QR{i}{r}{x}$}&\small{$ \bm{\bar{3}}_x$ }&\small{$\bm{1}$}&\small{$\bm{3}^r$}&\small{$\bm{3}^{i}$}\tabularnewline
\vspace{1mm}
\small{Left-Quark} \hfill \small{$\QL{i}{x}{l}$}&\small{$\bm{3}^x$ }&\small{$\bm{\bar{3}}_l$}&\small{$\bm{1}$}&\small{$\bm{3}^{i}$}\tabularnewline
\hline
\small{Colour-adjoint} \hfill \small{$\bm{\Delta}^a_{\mathrm C}$}&\small{$ \bm{8}^a$ }&\small{$\bm{1}$}&\small{$\bm{1}$}&\small{$\bm{1}$}\tabularnewline
\small{Left-adjoint}\hfill \small{$\bm{\Delta}^a_{\mathrm L}$}&\small{$ \bm{1}$ }&\small{$\bm{8}^a$}&\small{$\bm{1}$}&\small{$\bm{1}$}\tabularnewline
\small{Right-adjoint} \hfill \small{$\bm{\Delta}^a_{\mathrm R}$}&\small{$ \bm{1}$ }&\small{$\bm{1}$}&\small{$\bm{8}^a$}&\small{$\bm{1}$}\tabularnewline
\small{Family-adjoint} \hfill \small{$\bm{\Delta}^a_{\mathrm F}$}&\small{$ \bm{1}$ }&\small{$\bm{1}$}&\small{$\bm{1}$}&\small{$\bm{8}^a$}\tabularnewline
\hline
\end{tabular}
\caption[]
{The minimal chiral superfield content of the SUSY $[\SU{3}{}]^3 \times \SU{3}{F}\subset \mathrm{E}_8$ model [with global family $\SU{3}{F}$].}
\label{table:ChiralSuperE8}
\end{table}

The soft SUSY-breaking potential contains
\begin{eqnarray}
\begin{aligned}
\label{eq:G}
&V^{\rm G}_{\rm soft} = \Big\{ m^2_{\bm{27}} \tilde{L}\tilde{L}^\dagger + m^2_{\bm{78}} 
\tilde{\Delta}^{*a}_{\mathrm L} \tilde{\Delta}^a_{\mathrm L} + 
\Big[ b_{\bm{78}} \tilde{\Delta}^a_{\mathrm L} \tilde{\Delta}^a_{\mathrm L}  \\
& +\; d_{a b c} \big( A_{\bm{78}} \tilde{\Delta}^a_{\mathrm L} \tilde{\Delta}^b_{\mathrm L} \tilde{\Delta}^c_{\mathrm L} + 
C_{\bm{78}} \tilde{\Delta}^{*a}_{\mathrm L} \tilde{\Delta}^b_{\mathrm L} \tilde{\Delta}^c_{\mathrm L}\big)  \label{eq:Vsg} \\ 
& +\; A_{G} \tilde{\Delta}_{\mathrm L}^aT_{\mathrm L}^a\,\big(\tilde{L}^\dagger \tilde{L}+
\tilde{Q}_{\mathrm L}^\dagger \tilde{Q}_{\mathrm L}\big) + \mathrm{c.c.}\Big] \Big\} \ltimes \mathbb{Z}_3 \\
& +\; \Big[A_{\bm{27}} \varepsilon_{ijk}  \tilde{Q}_{\mathrm L}^i \tilde{Q}_{\mathrm R}^j \tilde{L}^k + \mathrm{c.c.}\Big]
\end{aligned}
\end{eqnarray}
accounting for gauge adjoint scalars $\tilde{\Delta}^a_{\mathrm{L,R,C}}$, and
\begin{eqnarray}
\begin{aligned}
\label{eq:Gl}
&V^{\rm Gl}_{\rm soft} = m^2_{\bm{1}} \tilde{\Delta}^{*a}_{\mathrm F} \tilde{\Delta}^a_{\mathrm F} + 
\big\{ b_{\bm{1}} \tilde{\Delta}^a_{\mathrm F} \tilde{\Delta}^a_{\mathrm F} + 
A_{\bm{1}} d_{a b c} \tilde{\Delta}^a_{\mathrm F} \tilde{\Delta}^b_{\mathrm F} \tilde{\Delta}^c_{\mathrm F} + \label{eq:Vsf} \\ 
&+\;A_{F} \tilde{\Delta}_{\mathrm F}^a\,\big(\tilde{L}^\dagger T_{\mathrm F}^a\tilde{L}\big) \ltimes \mathbb{Z}_3 + \mathrm{c.c.} \big\}
\end{aligned}
\end{eqnarray}
for interactions involving family octets $\tilde{\Delta}^a_{\mathrm F}$, where $T_{\mathrm A}^a$ are the $\SU{3}{A}$ generators such that 
$\Scale[0.9]{\tilde{L}^\dagger T_{\mathrm L}^a\tilde{L}\equiv \SLLRs{k}{l}{r} \TuL{a}{l'}{l}\SLLR{k}{l'}{r}}$ etc, and summation over 
$\mathbb{Z}_3$ permutations is implied by the symbol $\ltimes \mathbb{Z}_3$. For completeness, we also include soft SUSY-breaking 
interactions in the fermion sector,
\begin{eqnarray}
\label{eq:Lsoft}
\mathcal{L}^{\rm ferm}_{\rm soft} = \Big\{ -\dfrac{1}{2} M_0 \tilde{\lambda}^a_{\mathrm L} \tilde{\lambda}^a_{\mathrm L} - 
M^{\prime}_0 \tilde{\lambda}^a_{\mathrm L} \Delta^a_{\mathrm L} + \mathrm{h.c.} \Big\} \ltimes \mathbb{Z}_3 \,.
\end{eqnarray}
Here, besides the gaugino Majorana mass $M_0$, the symmetry allows a Dirac mass term parameterized by $M_0^{\prime}$.

Notably, by setting all the soft SUSY-breaking parameters to zero the model still allows for the trinification SSB with a T-GUT-breaking but SUSY-preserving 
stable vacuum giving rise to an effective SUSY LR-symmetric model below the GUT scale. At the moment, however, it is unclear if one could generate 
a consistent soft SUSY-breaking and gauge symmetry SSB in such an effective model providing a large splitting between the GUT and SM energy scales 
as required by phenomenology. We leave this open question to further studies taking into account the generic soft SUSY-breaking sector 
in the considered T-GUT as specified above.

In SUSY models with Dirac gauginos (such as minimal supersymmetric SM) the additional adjoint superfields spoil the gauge couplings' unification. This
problem is resolved in the so-called minimal Dirac-gaugino supersymmetric standard model \cite{Benakli:2014cia,Benakli:2016ybe}
inspired by $\SU{3}{}^3$ T-GUTs. In the SHUT model this problem is also resolved but, in a more elegant way, offering a framework that 
accommodates both the Dirac gauginos and the unified gauge coupling $g_{\mathrm{U}}$. Furthermore, the proton is stabilized to all orders 
in perturbation theory due to an accidental $\U{B}$. This global baryon symmetry is then preserved all the way down to the SM scale since 
none of the ($\tilde{Q}_{\mathrm L}$, $\tilde{Q}_{\mathrm R}$) squarks carrying the baryon number ($B=+1/3$, $-1/3$) acquire a \vev 
\cite{Camargo-Molina:2016bwm} (see also Ref.~\cite{Sayre:2006ma}).

\section{SUSY T-GUT symmetry breaking}

The presence of family $\SU{3}{F}$ symmetry together with adjoint superfields 
$\bm{\Delta}^a_{\mathrm{A}}$ allows for a consistent trinification SSB which is rather clean compared to older SUSY T-GUT realizations. It also provides, 
in particular, SM-like fermion candidates whose masses are protected 
from GUT-scale contributions. Choosing a VEV along the $\tilde{\Delta}^8_{\mathrm{A}}$ direction yields the rank-preserving trinification 
SSB
\begin{eqnarray}
\SU{3}{A}\to\SU{2}{A} \times \U{A} \,, \qquad \mathrm{A}=\mathrm{L},\mathrm{R},\mathrm{F}~.
\end{eqnarray}
Such a VEV choice is 
\begin{eqnarray}
\langle \tilde{\Delta}^8_{\mathrm L} \rangle \equiv v_{\mathrm L}\,, \qquad
\langle \tilde{\Delta}^8_{\mathrm R} \rangle \equiv v_{\mathrm R}\,, \qquad
\langle \tilde{\Delta}^8_{\mathrm F} \rangle \equiv v_{\mathrm F} 
\end{eqnarray}
(where $v_{\mathrm L}=v_{\mathrm R} \equiv v$ is required by vacuum stability) which provides the SSB scheme
\begin{eqnarray}
\nonumber
&& \left[ \SU{3}{C} \times \SU{3}{L} \times \SU{3}{R}\right] \ltimes \mathbb{Z}_3 \times \SU{3}{F} \\
&&  \quad \overset{v,\,v_{\mathrm F}}{\to} \quad \SU{3}{C}\times \big[ \SU{2}{L} \times \SU{2}{R}  \label{eq:brk}\\ 
&& \qquad \times\; \U{L} \times \U{R} \big] \ltimes \mathbb{Z}_2 \times \SU{2}{F} \times \U{F}\,,
\nonumber
\end{eqnarray}
in addition to implicit accidental symmetries such as $\U{B}$. Here, the square brackets denote parts gathered under the permutation 
symmetries. 

After the T-GUT symmetry breaking (\ref{eq:brk}) the fermionic tritriplets $L$, $Q_{\mathrm{L}}$, and ${Q}_{\mathrm{R}}$ are split into 
blocks revealing, e.g.,~massless $\SU{2}{L}$ [$\SU{2}{R}$] doublets of leptons $E_{\mathrm{L}}\equiv (e_{\mathrm{L}},\,\nu_{\mathrm{L}})$ 
[$E_{\mathrm{R}}\equiv (e_{\mathrm{R}}^c,\,\nu_{\mathrm{R}}^c)$] and quarks $q_{\mathrm{L}}\equiv (u_{\mathrm{L}},\,d_{\mathrm{L}})$ 
[$q_{\mathrm{R}}\equiv (u_{\mathrm{R}}^c,\,d_{\mathrm{R}}^c)$], whose first and second generations form $\SU{2}{F}$ doublets. Notably, 
the matching of Yukawa couplings in subsequent EFT scenarios is greatly simplified due to the unified Yukawa interactions in the considered T-GUT.

\section{Left-right-symmetric effective theory}

We have found that the high-scale SHUT model gives rise to a non-SUSY LR 
$\SU{2}{L} \times \SU{2}{R}$-symmetric EFT [Eq.~\ref{eq:brk}] as long as the quadratic and trilinear soft SUSY-breaking terms 
are small compared to the GUT scale. Here, we briefly discuss an important class of its characteristic low-energy scenarios where 
(i) all the gauge-adjoint $\tilde{\Delta}_{\mathrm{L},\mathrm{R},\mathrm{C}}$ and the flavour-adjoint $\tilde{\Delta}^{1,2,3,8}_{\mathrm{F}}$ scalars are heavy, 
thus are integrated out at the T-GUT-breaking (or, simply, GUT) scale, and (ii) the fundamental scalars $\tilde{L}$ 
are lighter than the GUT scale and are kept in the LR-symmetric EFT. This is indeed the most natural choice as the masses of the latter are solely governed by 
soft SUSY-breaking interactions while those of the former also contain large $\mathcal{F}$- and $\mathcal{D}$-term contributions of the order of the GUT scale. 
In particular, assuming for simplicity the superpotential and soft SUSY-breaking parameters to be real, it follows from Eqs.~\eqref{eq:G} and~\eqref{eq:Gl} 
that the masses of the scalar components of the tritriplets $\bm{L}$, $\bm{Q}_{\mathrm L}$, and $\bm{Q}_{\mathrm R}$ are of the form
\begin{align}
\label{eq:27scalars}
m^2_{\tilde{\varphi}_i} =m^2_{\bm{27}} + c_1^i A_{\rm G} v + c_2^i A_{\rm F} v_{\rm F} \,,
\end{align}
where the index $i$ runs over all fundamental scalars and $c_{1,2}^i$ are irrational constants. We can now relate all dimensionful parameters 
to the T-GUT-breaking \vev as $m^2_{\bm{27}} \equiv \alpha_{\bm{27}} v^2$, $A_{\rm G} \equiv \sigma_{\rm G} v$, 
$A_{\rm F} \equiv \sigma_{\rm F} v$, and $v_{\rm F} \equiv \beta v$.
Here, $\alpha_{\bm{27}},\,\sigma_{\rm G},\,\sigma_{\rm F}\ll 1$ are small, as they parametrize unknown details of soft SUSY breaking, while 
$\beta \sim \mathcal{O}\left( 1 \right)$ such that both gauge and family SSBs occur simultaneously. This allows us to recast the scalar masses as
\begin{align}
\label{eq:27recast}
m^2_{\tilde{\varphi}_i} = v^2  \left( \alpha_{\bm{27}} + c^i_1 \sigma_{\rm G} + c^i_2 \beta \sigma_{\rm F} \right) 
\equiv v^2 \omega_{\tilde{\varphi}_i} \,, \quad \omega_{\tilde{\varphi}_i} \ll 1 \,.
\end{align}
Interestingly, the light scalar spectrum of the effective LR-symmetric model is fully determined by three independent small parameters characterizing 
the soft SUSY-breaking sector and thus is protected from gaining the GUT-scale radiative corrections. 
Choosing, for example, $\Scale[0.9]{\omega_{\tilde{H}^{(3)}} \equiv \xi}$, $\Scale[0.9]{\omega_{\tilde{E}_{\rm L,R}^{(1,2)}} \equiv \delta}$, and 
$\Scale[0.9]{\omega_{\tilde{H}^{(1,2)}} \equiv \kappa}$, one obtains
\begin{equation}
\begin{aligned}
m_{\tilde{H}^{(3)}}^{2} 	&=	v^{2}\xi\,, \\
m_{\tilde{E}_{\rm L,R}^{(3)}}^{2}	&=	v^{2}\left(\delta+\xi-\kappa\right)\,, \\
m_{\tilde{\phi}^{(3)}}^{2}	&=	v^{2}\left(2\delta+\xi-2\kappa\right)\,,  \\
m_{\tilde{q}_{\rm L,R}^{(3)}}^{2}	&=	\tfrac{1}{3}v^{2}\left(\delta+3\xi-\kappa\right)\,, \\
m_{\tilde{D}_{\rm L,R}^{(3)}}^{2}	&=	\tfrac{1}{3}v^{2}\left(4\delta+3\xi-4\kappa\right)\,, \\
\end{aligned} \qquad
\begin{aligned}
m_{\tilde{H}^{(1,2)}}^{2}	&=	v^{2}\kappa\,, \\
m_{\tilde{E}_{\rm L,R}^{(1,2)}}^{2} 	&=	v^{2}\delta\,, \\
m_{\tilde{\phi}^{(1,2)}}^{2}	&=	v^{2}\left(2\delta-\kappa\right)\,, \\
m_{\tilde{q}_{\rm L,R}^{(1,2)}}^{2}	&=	\tfrac{1}{3}v^{2}\left(\delta+2\kappa\right)\,, \\
m_{\tilde{D}_{\rm L,R}^{(1,2)}}^{2}	&=	\tfrac{1}{3}v^{2}\left(4\delta-\kappa\right)\,.
\end{aligned}
\end{equation}
where $\xi$, $\delta$ and $\kappa$ determine all possible mass hierarchies in the scalar spectrum in 
the LR-symmetric EFT at the GUT scale.
Together with quartic, Yukawa, and gauge couplings, they control the initial conditions and shape 
of the RG flow and therefore define a particular SSB scheme affecting the features of the low-energy EFT limit. 
For example, setting $\kappa \ll \xi \ll \delta$ one finds that $m^2_{\tilde{H}^{(1,2)}} \ll m^2_{\tilde{H}^{(3)}} \ll 
m^2_{\rm others} \ll v^2$. One of the possible symmetry-breaking schemes down to the SM gauge group consists of 
two subsequent steps that can be induced by the \vevs $\langle\tilde{\phi}^{(3)}\rangle\equiv \langle\SLLR{3}{3}{3} \rangle$ 
and $\langle \tilde{\nu}_{\mathrm{R}}^{(2)}\rangle \equiv \langle\SLLR{2}{3}{1}\rangle$ at well-separated scales. 
This is represented by the following SSB chain:
\begin{eqnarray}
\begin{aligned} \label{eq:brk2}
&\SU{3}{C} \times [\SU{2}{L} \times \SU{2}{R} \times \mathrm{U}(1)_{\mathrm{L}} \times \mathrm{U}(1)_{\mathrm{R}}] \ltimes \mathbb{Z}_2 \quad   
\\
&\overset{\langle\tilde{\phi}^{(3)}\rangle}{\to} \quad \SU{3}{C} \times [\SU{2}{L} \times \SU{2}{R}] \ltimes \mathbb{Z}_2 \times \mathrm{U}(1)_{\mathrm{L}+\mathrm{R}}  \quad   
\\
&\overset{\langle \tilde{\nu}_{\mathrm{R}}^{(2)}\rangle}{\to} \quad \SU{3}{C}\times  \SU{2}{L} \times  \mathrm{U}(1)_{\mathrm{Y}} \,,
\end{aligned}
\end{eqnarray}
where only the gauge symmetry and $\mathbb{Z}_2$ are shown.

Consider the SSB chain (\ref{eq:brk2}) in more detail. Due to the presence of both Majorana and Dirac mass terms in the fermion-adjoint 
sector, with a large splitting one recovers light neutralino- and gluinolike states in the LR-symmetric EFT with masses
$m_{\mathcal{S}_{\rm L,R}}~\simeq~m_{\mathcal{T}_{\rm L,R}}~\simeq~m_{\tilde{g}}~\simeq~2 M_0$ in terms of 
the soft SUSY-breaking parameter $M_0~\ll~v~\sim~\mu_{\bm{78}}$. Here,
the $\SU{2}{L,R}$ triplet $\mathcal{T}_{\rm L,R}$ and singlet $\mathcal{S}_{\rm L,R}$ states emerge
from a decomposition of the $\SU{3}{L,R}$ octets as $\bm{8} \to {\bm{3}_0} \oplus {\bm{2}_1} \oplus 
{\bm{2}_{-1}} \oplus {\bm{1}_0}$, and $\tilde{g}$ is the lightest gluino. On the other hand, as long as 
$M_0~\sim~\langle\tilde{\phi}^{(3)}\rangle~\ll~v$, these gauginolike states will be integrated out at the
$\mathcal{O}\big(\langle\tilde{\phi}^{(3)}\rangle\big)$ scale. Thus, in the resulting $\SU{2}{L} \times \SU{2}{R}\times 
\mathrm{U}(1)_{\mathrm{L}+\mathrm{R}}$ EFT, the gaugino-lepton mass terms do not appear and 
the SM fermions are guaranteed to remain massless until the electroweak scale. Conveniently, the charges of 
the weak-singlet (non-SM) down-type quarks allow them to gain masses at the LR-breaking scales $\langle\tilde{\phi}^{(3)}\rangle$, 
$\langle \tilde{\nu}_{\mathrm{R}}^{(2)}\rangle$ via the high-scale Yukawa terms of the form $Q_{\mathrm L}Q_{\mathrm R}\tilde{L}$. 

\section{Significance, expectations and future work}

The proposed $\mathrm{E}_6$-extended SHUT model represents a promising way of unifying 
the light Higgs scalar and SM lepton sectors into the same supermultiplet $\bm{L}$, where [due to the trinification SSB via adjoint scalar \vevs and the family $\SU{3}{F}$] 
the SM fermions are protected from gaining masses in the high-scale model, in consistency with the SM. The inclusion of $\SU{3}{F}$ also results in the high-scale unification 
of the tree-level quark-lepton Yukawa couplings in the current framework [see $\lambda_{\bm{27}}$ in Eq.~\eqref{eq:WE8}]. Due to the emergent Yukawa and Higgs-lepton 
unification properties, the SHUT model has a relatively low number of free parameters at the GUT scale without introducing additional Higgs multiplets besides those 
in $\mathrm{E}_8$ and also without assuming any universality in the soft SUSY-breaking sector. While potentially sharing some of the key features 
of the non-SUSY T-GUT scenario discussed in Ref.~\cite{Camargo-Molina:2016bwm}, the SHUT model brings a straightforward explanation to some of its seemingly 
arbitrary characteristics such as the presence of scalars and fermions with the same quantum numbers. 

In particular, in Ref.~\cite{Camargo-Molina:2016bwm} it was demonstrated that in the non-SUSY T-GUT the LR symmetry breaking down to the SM gauge group can be 
initiated radiatively through the RG evolution. The circumstances under which the model leads to a realistic mass spectrum at lower energies were also explored, as well as 
aspects of its one-loop stability. Indeed, due to the running of a mass squared of a scalar $\SU{2}{R/F}$ bidoublet $(\tilde{e}_{i=1,2},\,\tilde{\nu}_{i=1,2})$ to a negative 
value at lower scales, the SSB can be triggered in the LR-symmetric EFT with a residual global $\SU{2}{F}$ down to the SM gauge symmetry [cf. the last SSB step in 
Eq.~(\ref{eq:brk2})]. Similar low-energy features could be present in the considered SHUT model as a plausible possibility, though they are not immediately guaranteed 
since its mass spectra differ from that of Ref.~\cite{Camargo-Molina:2016bwm}. A better understanding of the radiative symmetry breaking in the resulting 
LR-symmetric EFT which determines the structure of the SM-like theory at low energies should be the subject of future studies.

In the SM-like EFT, resulting from the chain (\ref{eq:brk2}), the three lightest SM Higgs $\SU{2}{L}$ doublets originating from the scalar $\SU{2}{L}\times 
\SU{2}{R} \times \SU{2}{F}$ tridoublet in the LR-symmetric EFT are expected to develop \vevs breaking the electroweak symmetry. As long as this property holds 
true, it provides a correct mass scale for the SM quarks in the second and third generations as well as gives rise to the Cabbibo mixing pattern at tree level. 
While there are no tree-level Higgsino, SM lepton, and first-generation quark masses in the high-scale theory, those can, in principle, be regenerated radiatively 
as soon as the LR and electroweak symmetries are broken. The EFT fermion mass spectra should thus be explored at least to one-loop order in following studies.

\section{Conclusions}

By unifying light Higgs bosons and SM leptons in the same supermultiplet of trinification, by breaking the trinification symmetry 
with adjoint scalar \vevs, and by introducing a global family symmetry, the SHUT model protects the SM fermions from gaining masses until the electroweak symmetry 
is broken while still ensuring the proton stability. The apparent simplicity of the SHUT model, originating from its gauge, Yukawa, and 
Higgs-lepton unification at the trinification breaking scale, makes it a very interesting candidate for further theoretical and phenomenological studies. 
Depending on the chosen symmetry-breaking scheme as well as on values of the high-scale couplings and the hierarchy between them, the path down to an effective 
SM-like theory could lead to vast and yet unexplored low-energy phenomena. While those are yet to be understood in full detail, the SHUT model presented here shows 
potential for reviving the trinification GUT model building. 

The first immediate task in further developments of the proposed high-scale SHUT model is to derive the basic properties of its SM-like EFT limit (at least, to one loop) 
and then to search for possible deviations from the characteristic SM signatures. This would allow us to set constraints on the SHUT parameter space and, possibly, to predict 
new smoking gun signals of new physics specific to the corresponding LR-symmetric EFT. The latter would then offer a plethora of opportunities for phenomenological 
studies of potentially observable beyond-SM phenomena in connection with the ongoing LHC and astroparticle physics searches.
\par
\noindent \textbf{\textit{Acknowledgments.}}~The authors would like to thank N.-E. Bomark, C. Herdeiro, W. Porod, and F. Staub for insightful discussions 
in the preliminary stages of this work. J. E. C.-M. was supported by the Crafoord Foundation. A. M. and M. S. are funded by 
Grants of the Funda\c{c}\~{a}o para a Ci\^{e}ncia e a Tecnologia (FCT), Portugal, No. SFRH/BPD/97126/2013 
and No. \mbox{SFRH/BPD/69971/2010} respectively.  R. P., A. O., and J. W. were partially supported by the Swedish Research Council, 
Contract No. 621-2013-428. The work in this paper is also supported by FCT funding to CIDMA (Center for Research and Development 
in Mathematics and Applications), Project No. UID/MAT/04106/2013. R. P. was partially supported by the Comisi\'{o}n Nacional de 
Investigaci\'{o}n Cient\'{\i}fica y Tecnol\'{o}gica (CONICYT) project No. PIA ACT1406.

\bibliography{bib}

\end{document}